\documentclass[conference]{IEEEtran}
\IEEEoverridecommandlockouts

\usepackage{cite}
\usepackage{amsmath,amssymb,amsfonts}
\usepackage{algorithmic}
\usepackage{graphicx}
\usepackage{textcomp}
\usepackage{xcolor}

\usepackage[utf8]{inputenc} 
\usepackage[T1]{fontenc}    
\usepackage{hyperref}       
\usepackage{url}  

\usepackage{booktabs}       
\usepackage{amsfonts}       
\usepackage{nicefrac}       
\usepackage{microtype}      
\usepackage{xcolor}         
\usepackage{graphicx}
\usepackage[square,sort,comma,numbers]{natbib}
\usepackage{doi}
\usepackage{amsmath}
\usepackage{subfig}
\usepackage[export]{adjustbox}
\usepackage{makecell, caption}
\usepackage{booktabs, siunitx, array,threeparttable}
\usepackage{multirow}
\usepackage{subcaption}

\def\BibTeX{{\rm B\kern-.05em{\sc i\kern-.025em b}\kern-.08em
    T\kern-.1667em\lower.7ex\hbox{E}\kern-.125emX}}

\begin{document}
\setcitestyle{square}
\title{Using RLHF to align speech enhancement approaches to mean-opinion quality scores\\
\thanks{This work was supported in part by the National Science Foundation under grant IIS-2235228, and in part by the Ohio Supercomputer Center.}
}

\author{\IEEEauthorblockN{Anurag Kumar}
\IEEEauthorblockA{
Computer Science \& Engineering \\
\textit{Ohio State University}\\
Columbus, OH, USA  \\
kumar.1109@osu.edu}
\and
\IEEEauthorblockN{Andrew Perrault}
\IEEEauthorblockA{
Computer Science \& Engineering \\
\textit{Ohio State University}\\
Columbus, OH, USA  \\
perrault.17@osu.edu}
\and
\IEEEauthorblockN{Donald S. Williamson}
\IEEEauthorblockA{
Computer Science \& Engineering \\
\textit{Ohio State University}\\
Columbus, OH, USA \\
williamson.413@osu.edu}\\
\and
}

\maketitle

\begin{abstract}
Objective speech quality measures are typically used to assess speech enhancement algorithms, but it has been shown that they are sub-optimal as learning objectives because they do not always align well with human subjective ratings. This misalignment often results in noticeable distortions and artifacts that cause speech enhancement to be ineffective. 
To address these issues, we propose a reinforcement learning from human feedback (RLHF) framework to fine-tune an existing speech enhancement approach by optimizing performance using a mean-opinion score (MOS)-based reward model. Our results show that the RLHF-finetuned model has the best performance across different benchmarks for both objective and MOS-based speech quality assessment metrics on the Voicebank+DEMAND dataset. Through ablation studies, we show that both policy gradient loss and supervised MSE loss are important for balanced optimization across the different metrics.
\end{abstract}


\section{Introduction}
In real-world environments, spoken communication is often obstructed by unwanted background noise that negatively affects perceptual quality and our ability to communicate effectively. Speech enhancement generally addresses this by using deep neural networks to remove unwanted noise. This is especially important to individuals with impaired hearing since they struggle to hear in noisy environments. 

Most speech enhancement approaches use deep learning to estimate a training target, where they are often trained using a mean-square error (MSE) loss objective. However, this is problematic because a higher MSE can result in better speech intelligibility for an enhanced signal \cite{8331910}. For this reason, researchers now use objective assessment measures, such as the Perceptual Evaluation of Speech Quality (PESQ) \citep{inproceedings_pesq}, Scale-Invariant Signal-to-Distortion Ratio (SI-SDR) \citep{LeRoux2018SDRH}, or Short-time Objective Intelligibility (STOI) \citep{5495701}, to assess performance. This, unfortunately, does not address the mismatch between the training objective (e.g., MSE) and the final evaluation metrics, where the two are not always in agreement. 

In recent years, generative adversarial networks (GANs) have shown improved performance and they address the mismatch problem by using objective measures for network optimization.  MetricGAN estimates time-frequency (TF) masks with a generator, while using a discriminator as a proxy function to estimate PESQ \cite{fu2019metricgan}, since 
PESQ is non-differentiable and cannot be used directly as a learning objective. Their goal is to optimize PESQ scores, where they outperform prior approaches according to PESQ. MetricGAN is further improved in \cite{fu2021metricgan}, where a replay buffer addresses catastrophic forgetting, a phenomenon common in GAN architectures. 
In \cite{Cao_2022}, a conformer-based GAN architecture (CMGAN) uses an additional MSE loss term for the generator network to optimize the real, imaginary, and magnitude components of the spectrograms, since these components address the phase response, which is important to perceptual quality \cite{7364200}. 
Unfortunately, using objective scores as learning objectives can be sub-optimal, because they do not always align with subjective scores, e.g., the mean-opinion score (MOS) or pairwise preference ratings from human listeners \citep{10.1109/TASLP.2019.2904850, 10.1109/TASLP.2021.3069302, inproceedings_santos, 7760550}. 

Subjective quality scores, like MOS, are standard measures of perceptual speech quality, so training objectives and subsequent performance evaluations should be based on subjective measures, especially in cases where human listeners will benefit most from speech enhancement. 
However, these scores involve expensive listening studies that require strict control of the listening environment. 
Fortunately, recent studies have produced subjective assessment data \citep{reddy2020interspeech, dong20_interspeech, yi22b_interspeech}, which can be used as part of a learning objective. Earlier attempts at using MOS scores include  \cite{nayem2023attentionbased} where authors developed a joint model for MOS estimation and speech enhancement.

In this paper, we better align the evaluation metrics and learning objectives by using estimated subjective quality scores as a reward signal in a Reinforcement Learning through Human Feedback (RLHF) framework, which is used to fine-tune existing enhancement approaches. A previous RL approach uses evaluation on a downstream automatic speech recognition task as a reward function to optimize the speech enhancement model \cite{shen2018reinforcement}. In \cite{7952122}, authors use relative PESQ improvements as a reward function to optimize a feed-forward network for speech enhancement. They argue that relative PESQ improvements are a better measure for the reward function as the absolute PESQ values are affected by outside parameters beyond the model's control. RLHF was initially introduced in \cite{10.1145/1597735.1597738} and improved in \cite{ouyang2022training, christiano2023deep} for natural language processing (NLP), where a reward model estimates human feedback with reasonable accuracy and better aligns NLP tasks with human assessment. We hypothesize that an RLHF framework will better align speech quality performance with human quality ratings, and it could eventually be used in an online manner to offer real-time enhancement, which is not possible with current approaches. Our reward model estimates MOS in a non-intrusive manner (i.e., without a clean reference), which has been done by prior approaches \citep{8937202, reddy2021dnsmos, Mittag_2021}. 
Our main contributions are that: (1) we provide a generalized RLHF training framework, one of the first attempts in the current literature that we know of, that uses RL to fine-tune an existing speech enhancement model with a MOS-based reward function, (2) through an ablation study, we show that fine-tuning using both a Proximal Policy Optimization (PPO) clip loss and supervised MSE loss is necessary for improvement across different objective and MOS-based speech quality metrics. 

\section{Methodology}
\label{sec:pagestyle}
\begin{figure}[!tb]
  \hspace{0.7cm}
  \includegraphics[scale=0.30]{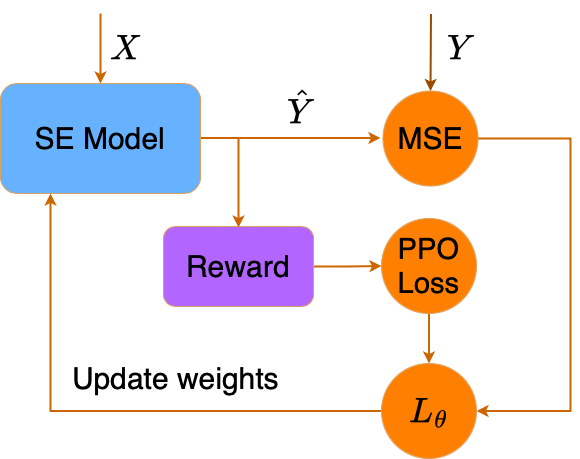}
  
  \captionsetup{justification=centering, margin=0cm}
  \caption{The proposed training framework. A reward is used to calculate the PPO clip loss, which is then combined with an MSE loss to update the SE model.} 
 \label{fig:1}
\end{figure}

In this section, we describe the conventional speech enhancement problem and the modifications required for fine-tuning using RLHF. An overview of the proposed RLHF approach is shown in Fig \ref{fig:1}.

\subsection{Speech Enhancement (SE)}
Given a noisy waveform $x \in \mathbb{R}^{l}$, we first extract the time-frequency (TF) representations, 
$X \in \mathbb{R}^{F \times T \times C}$, using the short-time Fourier transform. Here $F$, $T$, $C$ denote the frequency bins, time frames, and channels respectively, where channel refers to the real, imaginary, or magnitude spectrograms. 
Let $f_\theta(\cdot)$ represent a neural network with parameters $\theta$ that takes $X$ as input and estimates a set of TF masks, 
$\hat{M} \in \mathbb{R}^{F \times T \times C}$, i.e. $\hat{M} = f_\theta(X)$.
%
The components of the enhanced spectrogram, $\hat{Y} \in \mathbb{R}^{F \times T \times C}$, are obtained by applying channel-wise transformations $g_c(\cdot)$ to the input using the estimated masks, i.e., $\hat{Y} = g_c(\hat{M},X)$.  

Approaches that predict TF masks are most suitable for an RLHF framework. Therefore, we fine-tune two different yet, well-performing approaches to prove the generality of our RLHF fine-tuning framework. MetricGAN+ \cite{fu2021metricgan} is an LSTM-based network that only predicts magnitude masks, whereas CMGAN \cite{Cao_2022} is a conformer model that predicts both magnitude and complex masks for the spectrograms. The specific transformations for CMGAN and MetricGAN+ are given in \cite{Cao_2022} and \cite{fu2021metricgan} respectively.  

CMGAN has a 3 channel $\hat{Y}$, each channel representing the magnitude, real and imaginary values of the enhanced spectrogram, whereas outputs from MetricGAN+ are a single channel enhanced magnitude spectrogram. The enhanced waveform $\hat{y} \in \mathbb{R}^{l}$ is obtained using an inverse TF transformation to put the enhanced TF representations back in the time domain.

\subsection{Speech Enhancement as an RL problem}
We define our supervised finetuning policy $\pi_{\theta}^{SFT}: S \rightarrow A$ to be the mapping $f_\theta(\cdot)$ parameterized by $\theta$. The state space, $S$, is continuous as it covers the distribution of all possible noisy spectrogram inputs. 
We interpret the set of masks, $\hat{M}$, predicted by the supervised policy as actions. Our action space, $A$, should cover the distribution of these masks and it is also continuous. An episode is a single-step (utterance level) operation where an enhanced signal is obtained from the policy. We modify the output of $f_\theta(\cdot)$ by adding zero-mean fixed-variance Gaussian noise $n \sim \mathcal{N}(0, \sigma)$ to get $\hat{M}^{RL} = \{\hat{m}_c^{RL}\}_{c=0}^{C-1}$, which is used to calculate the enhanced TF representation, $\hat{Y}$. This enables us to calculate the likelihood of our action and implement policy gradient methods. We refer to the set of predicted masks, $\hat{M}^{RL}$, as $a$ (an action).  We denote the policy optimized by RLHF as $\pi^{RL}_\theta$.
\begin{equation}
\label{eq:m_rl}
\begin{gathered}
    \hat{M^{RL}} = f_\theta([X_0, X_1, .. X_{c-1}]) + n
\end{gathered}
\end{equation}

\subsection{Our RL framework}
To incorporate human feedback, we use the NISQA model \cite{Mittag_2021}, a commonly-used predictor of MOS, as part of our reward model. Let the NISQA reward model be referred to by $r_{\phi}(\cdot)$, where $\phi$ are the NISQA model parameters. 
 Following the intuition in \cite{7952122}, we use the relative improvement in NISQA MOS between the outputs from $\pi_{\theta}^{SFT}$ and $\pi_{\theta}^{RL}$ as our reward signal, $r_{mos}$. We also experiment with an objective metric reward signal, $r_{pesq}$, to observe the difference in metrics optimized by either method. Here $y$ refers to the reference clean speech waveform. Let $\hat{y}_{rl}$ be the estimated speech waveform from $\pi_{\theta}^{RL}$ and $\hat{y}_{sft}$ be the estimated speech waveform from $\pi_{\theta}^{SFT}$. Then,
\begin{equation}
\label{eq:rew}
\begin{gathered}
    r_{mos} = r_{\phi}(\hat{y}_{rl}) - r_{\phi}(\hat{y}_{sft}) \\
    r_{pesq} = PESQ(\hat{y}_{rl}, y) - PESQ(\hat{y}_{sft}, y) 
\end{gathered}
\end{equation}

The overall objective maximized by our environment is given by $J_{\theta}$
\begin{equation}
     J_{\theta} =  \bigg[r - \beta \times \text{KL}\bigg(\pi_{\theta}^{RL}(\hat{y}|x),\pi_{\theta}^{SFT}(\hat{y}|x)\bigg)\bigg] 
\end{equation}
%
Here, $r$ refers to the different reward signals in \eqref{eq:rew} and $\beta$ controls the KL divergence between our fine-tuned policy and the pre-trained policy. We refer to the optimized policy before and after the optimization step as $\pi^{RL}_{\theta_{old}}$ and $\pi^{RL}_{\theta}$, respectively. The subscript $t$ denotes the respective values associated with the sampled noisy signal batch $x_t$. We modified the clipped loss objective of PPO by replacing the advantage function $A_t$ with $J_{\theta_t}$, 
which allows us to avoid using a critic network, simplifying the training procedure. As the episodes are a single step and the reward definition already includes an appropriate reward baseline (the performance of the SFT policy), a critic network is unnecessary.
\begin{equation}
\begin{split}
    L_{ppo-clip}& (\theta) = \\ 
                   & \mathbb{E}_{x_{t} \sim D_{SFT}} 
                    \bigg[
                    \text{min}\bigg(\frac{\pi^{RL}_{\theta}(\hat{y_t}|x_t)}{\pi^{RL}_{\theta_{old}}(\hat{y_t}|x_t)}J_{\theta_t} , \\
                  & \text{clip}\bigg(\frac{\pi^{RL}_{\theta}(\hat{y_t}|x_t)} {\pi^{RL}_{\theta_{old}}(\hat{y_t}|x_t)}, 1-\epsilon, 1+\epsilon\bigg)J_{\theta_t} \bigg)
                    \bigg]
    \label{eq:ppo}
\end{split}
\end{equation}

Since PPO is an on-policy optimization method, we used $\pi^{RL}_{\theta_{old}}$ to execute an episode and store the associated tuple $(x_t, a_t, J_{\theta_t}, \pi^{RL}_{\theta_{old}}(a_t|x_t))$ in an experience buffer. The values from the experience buffer are extracted to calculate the clip loss in \eqref {eq:ppo}. $\epsilon$ controls the step size for the PPO loss update. Here $D_{SFT}$ refers to the dataset used for supervised fine-tuning to get the initial pre-trained policy $\pi_{\theta}^{SFT}$. We also mix pretraining gradients during the finetuning stage. Apart from using the KL penalty in the objective function $J_\theta$, the pre-trained loss was included to keep our $\pi_{\theta}^{RL}$  parameter distribution from diverging too much from the $\pi_{\theta}^{SFT}$ parameter distribution. 
The overall loss objective that updates $\theta$ is given by $L_\theta$, 
\begin{equation}
\label{eq:ovl}
    L_\theta = L_{ppo-clip} + \lambda L_{MSE}
\end{equation}
We use $\lambda$ to control the weight of pre-training gradients on our final outputs.
$L_{MSE}$ is a weighted sum of mean-square errors across channels of the predicted and reference spectrograms. 
\begin{equation}
\label{eq:mse}
\begin{gathered}
  L_{MSE} = 
  \begin{cases}
    L_0, & C = 1 \\
    \alpha L_0 + (1-\alpha)\sum_{c=1}^{C-1}{L_c}, & C > 1 \\
  \end{cases}
\end{gathered}
\end{equation}
In the above eq, $L_c$ is defined as $(Y_c - \hat{Y}_c)^2$. $\alpha$ controls the weight of the two terms on the overall MSE loss.

\section{Experimental Setup}
\subsection{Dataset}

We evaluate our proposed methodology on the commonly used, publicly available VoiceBank+DEMAND dataset \cite{6709856}. The training set contains 11572 utterances from 28 different speakers and the testing set has 824 utterances from 2 unseen speakers. 
Furthermore, we use the test set of the LibriMix dataset \citep{cosentino2020librimix} to evaluate our model's estimated MOS quality scores. Given the lack of speaker diversity in the VCTK dataset, the test set of LibriMix contains 40 different speakers with up to 3 speaker noisy samples and 3000 samples for each speaker count. We use all 9000 signals for evaluation. 
 All utterances are sampled at 16kHz for our experiments. For all feature extraction, we use a 25 ms Hanning window with an overlap of 6.25 ms for CMGAN and a 32 ms Hanning window with an overlap of 16 ms for MetricGAN+. These are the original training parameters for the respective models.   

\subsection{Training}

We use a batch size of 4 and accumulated gradients for 16 steps for CMGAN and a batch size of 8 and accumulated gradients for 8 steps to fine-tune MetricGAN+, each resulting in an overall batch size of 64. The remaining architectural and pre-processing details are as followed in the original papers. For PPO, we set $\epsilon$ to 0.01. The values for $\alpha, \beta$, and $\lambda$ are set to 0.7, 0.0001, and 1, respectively. The $\sigma$ of the Gaussian noise is set to 0.01. While any larger amount of noise simply worsened the outputs, we did not see a substantial difference among the outputs generated from masks with noise having a smaller $\sigma$. Thus, $\sigma$ of 0.01 allowed maximum variance during the RL fine-tuning.
%
The learning rate for all our experiments is set to $1e^{-6}$. 
Our fine-tuned models are referred to as CMGAN\_PPO and MetricGAN\_PPO. All our experiments were run on a single NVIDIA A100 GPU.  

\subsection{Baselines}
We present multiple baselines reported within the last few years. Baselines can be categorized into two groups: (1) baselines that predict the time domain signal, 
DEMUCS \cite{defossez2020real} and SE-Conformer \cite{kim21h_interspeech}
, and (2) baselines that predict time-frequency (TF) representations, 
PFPL \cite{inproceedings}, MetricGAN+, DB-AIAT \cite{9746273}, DPT-FSNet \cite{9746171}, CMGAN and MP-SENet \cite{mpsenet}. 

We either report the results directly from the original paper or reproduce the outputs from inferencing using released checkpoints. The ones that were reproduced are reported with the (repr) tag. We chose to reproduce the results of MP-SENet, CMGAN, and MetricGAN+ as they had publicly available source code and checkpoints. They all predict TF masks and have among the best-reported results on the Voicebank+DEMAND dataset.

\begin{table}[!b]
\centering
\captionsetup{justification=centering, margin=0cm}
\caption{NISQA MOS scores between the pre-trained CMGAN baseline and CMGAN\_PPO using different datasets. We report the NISQA MOS scores on VCTK, LibriMix single, and 2-speaker and 3-speaker audio sets.}
\label{tab:2}
\begin{tabular}{lcccc}
\toprule
\multirow{2}{*}{Model} & \multirow{2}{*}{Voicebank} & \multicolumn{3}{c}{LibriMix}            \\ \cline{3-5} 

\multicolumn{1}{c}{} &  & 1 Spk & 2 Spk & 3 Spk \\\midrule
Noisy      & 3.03              & 1.52           & 1.47           & 1.37           \\ \hline
\\
MetricGAN+ & 3.99 & 2.62 & \textbf{2.47} & 2.32 \\ 
MetricGAN\_PPO & \textbf{4.08} & \textbf{2.65} & 2.46 & \textbf{2.35} \\\hline \\
CMGAN      & 4.65              & 4.12           & 3.87           & 3.57           \\
CMGAN\_PPO & \textbf{4.73}              & \textbf{4.17}           & \textbf{3.93}           & \textbf{3.62} \\\bottomrule
\end{tabular}
\end{table}

\section{Results \& Discussion}
We evaluate our models on a set of commonly used speech quality metrics. For estimated subjective scores, we evaluate the model's performance using the NISQA MOS ratings. These ratings have a range of [1,5]. For the objective metrics, we use PESQ with a range of [-0.5, 4.5], segmental signal-to-noise ratio (SSNR) and SI-SDR, to measure the model's performance.

\begin{figure}[t]
  \vspace{-0.5cm}
  \hspace{-1.4cm}
  \includegraphics[scale=0.44]
  {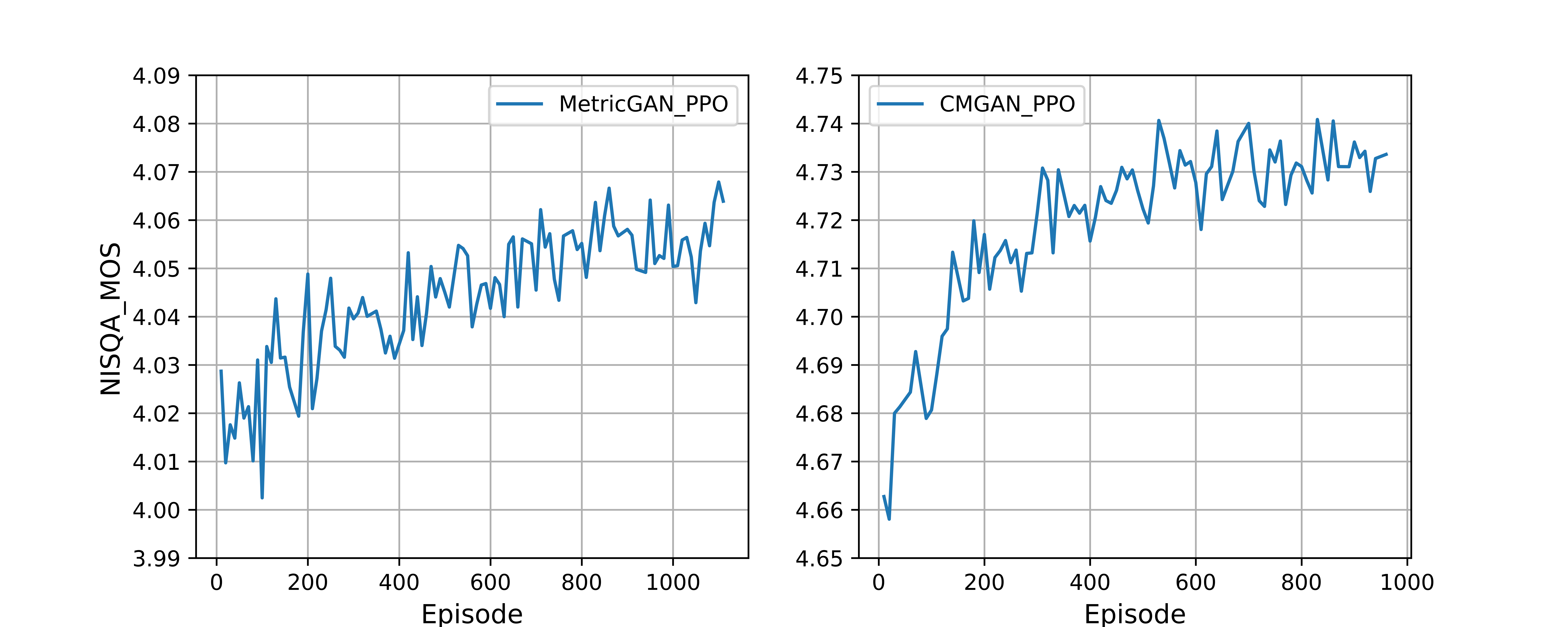}  
  \captionsetup{justification=centering, margin=0cm}
  \caption{NISQA MOS on VCTK test set recorded every 10 episodes during fine-tuning using $r_{mos}$ as reward function.} 
 \label{fig:2}
\end{figure}

We evaluate the baselines and our proposed fine-tuned models on an unseen multi-speaker dataset LibriMix and the seen VCTK test set. Table \ref{tab:2} shows the NISQA MOS scores of the enhanced signals for the pre-trained and proposed models. While both MetricGAN\_PPO and CMGAN\_PPO consistently increase the NISQA MOS scores on the VCTK dataset, we observe that CMGAN and CMGAN\_PPO perform consistently well across the unseen dataset LibriMix for all speaker counts unlike MetricGAN+. We conducted a t-test to measure the statistical significance of these results. We obtained p-values of 4.07e-05, 2.59e-03, 4.48e-04 and 2.51e-03 for the NISQA\_MOS scores on the VCTK, LibriMix(1spk), LibriMix(2spk) and LibriMix(3spk), respectively. The comparison results for objective metrics on the VCTK test set are shown in Table~\ref{tab:1}. Our proposed models show superior performance across all of the benchmarks without significant changes to PESQ when compared to their respective baselines. We see consistent improvements in SSNR and SI-SDR after fine-tuning (MetricGAN\_PPO and CMGAN\_PPO) compared to their respective baselines MetricGAN+ and CMGAN. The observed improvement in SI-SDR, SSNR, and NISQA MOS scores while observing negative or no change to PESQ supports the general effectiveness of our proposed method. While our main objective is to improve quality and not intelligibility, our method does not degrade STOI which largely remains consistent. Fig \ref{fig:2} shows the increasing trend of the aggregated NISQA MOS scores of the VCTK test set recorded every 10 episodes throughout fine-tuning. 

To study the effect of a policy gradient loss, we also conduct an ablation study with different reward types. For CMGAN+MSE, CMGAN is further fine-tuned with the MSE loss described in Eq.~\eqref{eq:mse}. The remaining CMGAN\_PPO models are fine-tuned using the objective shown in Eq.~\eqref{eq:ovl} and different rewards, see Table \ref{tab:3}. Further training of CMGAN with a MSE loss results in inferior scores compared to those obtained from approaches fine-tuned with both the policy gradient and MSE loss. This means that the PPO loss improves performance. We observe that reward signals $r_{mos}$ and $r_{pesq}$ individually optimize a different subset of objective metrics. We get the highest PESQ score with $r_{pesq}$, whereas the highest NISQA\_MOS, SSNR and SI-SDR scores are obtained when finetuning with $r_{mos}$. We also trained a model with a combined reward function, $r_{comb} = r_{mos}+r_{pesq}$, but found subpar performance on all metrics compared to the models trained using either of the two reward functions individually. 

In our experiment with further MSE fine-tuning, the performance on evaluation metrics saturated quickly within the first epoch. The RLHF framework gives us an alternate way to calculate gradients that maximize the likelihood of outputs with greater rewards and thus introduces newer gradients. However, this can lead to poor generalization capacity. Using both MSE and policy gradient loss helps with generalization and larger gradient variance. This is shown through better improvements from CMGAN\_PPO across MOS-based and objective evaluation metrics as opposed to CMGAN+MSE finetuning.  

\noindent
\begin{table}[!t]
\centering
\captionsetup{justification=centering, margin=0cm}
\caption{This table compares baselines with our proposed approach on the Voicebank+DEMAND test set. We put "-" in place for metrics not reported in the original papers.}
\label{tab:1}
\begin{threeparttable}

\begin{tabular}{@{} lcccccc @{}} 
\toprule
{Model} & {STOI} & {PESQ} & {SSNR} & {SI-SDR}\\ 

\midrule
  Noisy & & 1.97 & 1.68 & 8.44 \\
  \hline\\
  DEMUCS & 0.95 & 3.07 & - & - \\ 
  PFPL & 0.95 & 3.15 & - & - \\
  SE-Conformer & 0.95 & 3.13 & - & - \\
  DB-AIAT & 0.96 & 3.31 & 10.79 & - \\
  DPT-FSNet & 0.96 & 3.33 & - & - \\
  MP-SENet (repr) & 0.96 & 2.95 & 9.91 & 19.63 \\ \hline
  MetricGAN+(repr) & 0.92 & \textbf{3.15} & 3.11 & 7.98 \\
  MetricGAN\_PPO (ours) & \textbf{0.93} & 3.12 & \textbf{3.51} & \textbf{9.57} \\ \hline
  CMGAN (repr.) & \textbf{0.96} & \textbf{3.39} & 10.35 & 20.02 \\
  CMGAN\_PPO (ours) & \textbf{0.96} & \textbf{3.39} & \textbf{10.93} & \textbf{20.28} \\ 
\bottomrule
\end{tabular}
\end{threeparttable}
\end{table}


 
\begin{table}[!b]
\centering
\footnotesize
\captionsetup{justification=centering, margin=0cm}
\caption{Ablation study for investigating the effect of different policy gradients and reward signals on performance. The results are obtained from the VCTK test set. N\_MOS is used as an abbreviation for NISQA MOS scores.}
\label{tab:3}
\begin{tabular}{@{} l cccccc @{}}
\toprule
{Model} & Reward & {PESQ} & {SSNR} & {SI-SDR} & N\_MOS \\

\midrule
  CMGAN (baseline) & - & 3.39 & 10.35 & 20.02 & 4.65 \\
  CMGAN+MSE & - & 3.41 & 10.74 & 20.14 & 4.66 \\ 
  CMGAN\_PPO  & $r_{pesq}$ & \textbf{3.44} & 10.89 & 20.23 & 4.72\\ 
  CMGAN\_PPO  & $r_{mos}$ & 3.39 & \textbf{10.93} & \textbf{20.28} & \textbf{4.73}\\
  CMGAN\_PPO  & $r_{comb}$ & 3.43 & 10.91 & 20.25 & 4.71\\
 
\bottomrule
\end{tabular}
\end{table}

\section{Conclusion}
We present an RLHF fine-tuning framework for speech enhancement. 
We also conduct an ablation study to quantify how a policy gradient affects performance and to assess the effect that reward functions have on policy gradient optimization. Through objective and MOS-based evaluations, we show our proposed model's improved performance over baselines on multiple datasets.

\bibliographystyle{IEEEtranS}
\bibliography{IEEEabrv,ref}

\end{document}